\documentclass[fleqn,usenatbib,useAMS]{mnras}
\usepackage{amssymb,amsmath}
\usepackage{graphicx,epstopdf}
\usepackage{soul,xcolor}

\usepackage{etoolbox}
\makeatletter
\patchcmd\@combinedblfloats{\box\@outputbox}{\unvbox\@outputbox}{}
    {\errmessage{\noexpand\@combinedblfloats could not be patched}}
\makeatother

\newcommand{\m}[1]{\boldsymbol{#1}}
\newcommand{\bra}[1]{\left#1}
\newcommand{\ket}[1]{\vphantom{\sqrt{0}}\right#1}

\title[Crust--magnetosphere coupling during evolution]{Crust--magnetosphere coupling during magnetar evolution and implications for the surface temperature}
\author[T. Akg\"{u}n et al.]
{T.~Akg\"{u}n$^1$\thanks{E-mail: akgun@astro.cornell.edu}, P.~Cerd\'{a}--Dur\'{a}n$^2$, J.A.~Miralles$^1$, and J.A.~Pons$^1$ 
\\$^1$Departament de F\'{i}sica Aplicada, Universitat d'Alacant, Ap. Correus 99, 03080 Alacant, Spain
\\$^2$Departament d'Astronomia i Astrof\'{i}sica, Universitat de Val\`{e}ncia, Dr. Moliner 50, 46100, Burjassot, Val\`{e}ncia, Spain}

\begin{document}
\label{firstpage}
\pagerange{\pageref{firstpage}--\pageref{lastpage}}
\maketitle

\begin{abstract}
We study the coupling of the force-free magnetosphere to the long-term internal evolution of a magnetar. We allow the relation between the poloidal and toroidal stream functions --- that characterizes the magnetosphere --- to evolve freely without constraining its particular form. We find that, on time-scales of the order of kyr, the energy stored in the magnetosphere gradually increases, as the toroidal region grows and the field lines expand outwards. This continues until a critical point is reached beyond which force-free solutions for the magnetosphere can no longer be constructed, likely leading to some large-scale magnetospheric reorganization. The energy budget available for such events can be as high as several $10^{45}\,$erg for fields of $10^{14}\,$G. Subsequently, starting from the new initial conditions, the evolution proceeds in a similar manner. The time-scale to reach the critical point scales inversely with the magnetic field amplitude. Allowing currents to pass through the last few meters below the surface, where the magnetic diffusivity is orders of magnitude larger than in the crust, should give rise to a considerable amount of energy deposition through Joule heating. We estimate that the effective surface temperature could increase locally from $\sim 0.1\,$keV to $\sim 0.3 - 0.6\,$keV, in good agreement with observations. Similarly, the power input from the interior into the magnetosphere could be as high as $10^{35} - 10^{36}\,$erg/s, which is consistent with peak luminosities observed during magnetar outbursts. Therefore, a detailed treatment of currents flowing through the envelope may be needed to explain the thermal properties of magnetars.
\end{abstract}


% Start of text.
% Sections 1 & 2: Introduction & Overview.
% Specify root document.
% !TEX root = paper.tex

\section{Introduction}
Magnetars display a wealth of distinctive highly energetic transient events, including recurrent short duration bursts, long duration outbursts accompanied by extended X-ray emission lasting several years, and giant flares \citep{2015SSRv..191..315M, 2017ARA&A..55..261K}. This activity is linked to the presence of strong magnetic fields, typically exceeding $10^{14}\,$G, and slowly evolving due to the Hall drift and Ohmic dissipation in the crust \citep{Jones1988,1992ApJ...395..250G,Pons:2009,Gourgouliatos:2016}. Typical magnetar temperatures in quiescence are $\sim 0.2-0.3\,$keV, with an emitting area of $\sim 1\,$km$^2$. During an outburst, the peak temperature can be several times higher, gradually recovering the quiescence state, or sometimes even a somewhat higher value \citep{2011ASSP...21..247R, 2018MNRAS.474..961C}. The high temperatures must be  maintained by some mechanism involving rapid dissipation of the magnetic field in a localized region, but the details are not fully understood.

Whatever the eventual triggering mechanism of these violent events, they are thought to occur as a result of the gradual build-up of energy, helicity and twist in the magnetosphere driven by the long-term evolution of the internal magnetic field \citep{Thompson:1996,Perna:2011,2014ApJ...794L..24B,2017ApJ...841...54T}. All models coincide in the expectation that when a sufficiently large twist is reached (typically $\gtrsim 1$\,rad) some kind of large-scale reorganization of the field structure must take place in an extremely short time (of the order of the Alfv\'{e}n time-scale) \citep{Lyutikov:2003,Gill:2010,2012ApJ...754L..12P,2013ApJ...774...92P,2017MNRAS.472.3914A,2017ApJ...844..133C}. Studying the equilibrium, stability and evolution of the magnetosphere is pivotal in understanding the processes that give rise to the X-ray activity of magnetars.

The equilibrium structure of a non-rotating axisymmetric magnetosphere is given by the Grad--Shafranov equation, which in general requires numerical solutions \citep[see, for example][]{2014MNRAS.445.2777F,2014MNRAS.437....2G,2015MNRAS.447.2821P,2016MNRAS.462.1894A,2018MNRAS.474..625A,2017MNRAS.468.2011K,2018MNRAS.477.3530K,2018MNRAS.475.5290K}. In \cite{2016MNRAS.462.1894A} and \cite{2018MNRAS.474..625A}, we constructed magnetospheric models with toroidal fields confined within a magnetic surface in the vicinity of the equator and smoothly joining to vacuum fields at large distances. We found that a force-free magnetosphere is able to store more energy than the vacuum (current-free) one, in some cases reaching up to $\sim 80\%$ more energy. However, almost invariably, the largest values of these energies correspond to configurations with field lines disconnected from the surface, which would likely be unstable as also argued by \cite{2018MNRAS.475.5290K} and \cite{2018MNRAS.477.3530K}. Therefore, such magnetospheres may not be realizable under normal conditions in magnetars. We also showed that for nearly all cases with disconnected field lines, lower energy configurations exist for the same parameters of the toroidal field, with field lines connected to the interior. In other words, the solutions of the Grad--Shafranov equation are degenerate, and the lower energy solutions correspond to the likely stable configurations. The maximum energy stored in such magnetospheres represents a moderate $\sim 25\%$ increase with respect to the vacuum case. This excess defines the energy budget available in the event of fast, global magnetospheric reorganizations of the field structure such as those associated with magnetar flares.

The energy stored in the magnetosphere (from the stellar surface all the way up to infinity) for a vacuum dipole field with an amplitude $B_{\rm pole}$ at the pole and a stellar radius $R_\star$ is 
\begin{equation}
	E_{\rm vac} = \frac{B_{\rm pole}^2 R_\star^3}{12} \approx 8.33 \times 10^{44} B_{14}^2 R_6^3 \ {\rm erg.}
\end{equation}
Here, $B_{14} = B_{\rm pole} / 10^{14}\,{\rm G}$ and $R_6 = R_\star / 10^6\,{\rm cm}$. Thus, for typical magnetar field strengths of the order of $10^{14}\,$G, the excess energy stored in the magnetosphere would be of the order of a few $10^{44}\,{\rm erg}$, consistent with observations of energetic events in magnetars.

We noted that lower energy (connected) field configurations are possible up to a maximum twist of $\varphi_{\rm max}\sim 1.5$\,rad, in agreement with other authors \citep{1994ApJ...430..898M,2002ApJ...574..332T,2012ApJ...754L..12P,2013ApJ...774...92P,2017MNRAS.468.2011K}. A significant fraction of the polar cap flux is already open when $\varphi \approx 2$, and it is reasonable to expect that increasing the twist further would lead to the sudden disruption of magnetospheric loops.

In \cite{2017MNRAS.472.3914A}, we investigated the coupling of such magnetospheric models to the long-term evolution of the interior computed by the code described in \cite{2012CoPhC.183.2042V}. We found that the magnetospheric currents can be maintained on time-scales of the order of hundreds or thousands of years depending on the field amplitude, while the energy stored in the magnetosphere gradually increases (as well as helicity and twist). This continues up to a \emph{critical point} beyond which no realistic force-free solutions can be constructed for the magnetosphere. At this point, we conjecture that some large-scale magnetospheric rearrangement must occur, releasing a large fraction of the stored energy. Subsequently, the quasi-steady evolution should proceed in a similar way from the new starting conditions. We also found that the spindown rate increases due to the gradual enhancement of the effective surface dipole strength, resulting in a braking index of $n < 3$ for most part of the evolution, consistent with measurements for pulsars and estimates for magnetars \citep{2015MNRAS.446..857L,2017MNRAS.466..147E}.

In this paper, we aim at understanding in greater detail how energy is transferred from the neutron star crust to the exterior, depending on the initial structure of the magnetic field, and to what extent such a transfer can proceed while maintaining force-free (but not current-free) magnetospheric equilibrium before some global reorganization (a burst or a flare) becomes inevitable. In contrast to \cite{2017MNRAS.472.3914A}, where we determined the dependence between the toroidal and poloidal stream functions in the magnetosphere through a best fit using a prescribed functional form with several free parameters, here we allow for a greater degree of freedom by considering the symmetric and antisymmetric modes driven by the internal evolution. We are primarily concerned with magnetars, where rotation can be safely neglected as their periods are relatively long (typically of the order of $10$\,s), with corresponding light cylinder radii of over $10^5$\,km --- well beyond the region of interest of a few stellar radii ($\lesssim 100$\,km).

The structure of this paper is as follows: in \S\ref{section_overview} we give a technical overview of our model; in \S\ref{section_results} we present sample simulations for the external field evolution coupled to the interior; in \S\ref{section_temperature} we consider the likely effects of our force-free model on the surface temperature; and in \S\ref{section_conclusions} we discuss the implications of our results.

\section{Technical overview}\label{section_overview}
\subsection{Internal evolution}
In the neutron star crust, the magnetic field evolution is given by the induction equation,
\begin{equation}
	\begin{split}
		\partial_t \m{B} & = - c \m\nabla \times \m{E} \\
		& = - \m\nabla \times \bra{[} f_H (\m\nabla \times \m{B}) \times \m{B} + \eta \m\nabla \times \m{B} \ket{]} \ .
	\end{split}
\end{equation}
The two terms correspond to the Hall effect and Ohmic dissipation, respectively. The Hall coefficient is defined as $f_H = c/4\pi e n_{\rm e}$, where $n_{\rm e}$ is the electron number density and $e$ is the elementary charge, and $\eta$ is the magnetic diffusivity and is related to the electrical conductivity $\sigma$ through $\eta = c^2/4\pi\sigma$.

In this work, we do not consider the evolution in the stellar core, which is dominated by the highly non-linear ambipolar diffusion, and is further complicated by the presence of neutron superfluidity and proton superconductivity \citep{1992ApJ...395..250G,2017MNRAS.471..507C,2017MNRAS.465.3416P}. We use the numerical code presented in \cite{2012CoPhC.183.2042V} to model the evolution in the crust. Relevant time-scales and observational implications are discussed in greater detail in \cite{2013MNRAS.434..123V}. Throughout this work, we use a neutron star model of mass $M_\star = 1.4 M_\odot$ and radius $R_\star = 11.6$\,km.

Most previous works on magnetic field evolution \citep[e.g.][]{2012CoPhC.183.2042V,2013MNRAS.434..123V} employ vacuum boundary conditions at the surface, where, given the radial component of the magnetic field ($B_r$), the tangential component ($B_\theta$) is calculated consistently with the boundary condition. In this work, as in \cite{2017MNRAS.472.3914A}, we generalize this to allow for the presence of currents and twist in the magnetosphere, while still neglecting the pressure and inertia of the plasma. The magnetosphere is assumed to adjust instantaneously to a new equilibrium at each time step, rapidly dissipating any transient perturbations.

\subsection{Magnetosphere}
An axisymmetric magnetic field can be represented in terms of the poloidal and toroidal stream functions ($P$ and $T$, respectively) or, alternatively, in terms of the azimuthal ($\phi$) components of the vector potential $\m{A}$ and the magnetic field $\m{B}$ as
\begin{equation}
	\begin{split}
	\m{B} & = \m\nabla P \times \m\nabla\phi + T \m\nabla\phi \\
	& = \m\nabla \times ( A_\phi \m{\hat{\phi}} )+ B_\phi \m{\hat{\phi}} \ ,
	\end{split}
	\label{field}
\end{equation}
in spherical coordinates $(r,\theta,\phi)$. Note that $P = A_\phi r\sin\theta$ and $T = B_\phi r\sin\theta$. Magnetic field lines are contours of constant $P$, with $P=0$ corresponding to the magnetic axis. In a static axisymmetric fluid, the Lorentz force cannot have an azimuthal component, implying that $T$ must be a function of $P$. The equilibrium structure  of a force-free magnetosphere is described by the corresponding Grad--Shafranov equation \citep[see][and references therein]{2016MNRAS.462.1894A},
\begin{equation}
	\triangle_{\rm GS} P + TT' = 0 \ .
	\label{GS_equation}
\end{equation}
Here, the prime denotes derivative with respect to $P$, and the Grad--Shafranov operator is given through
\begin{equation}
	\triangle_{\rm GS} = \partial_r^2 + \frac{1 - \mu^2}{r^2} \partial_\mu^2 \ ,
	\label{GS_operator}
\end{equation}
where $\mu = \cos\theta$. Current-free further requires $T=0$. The force-free condition implies that the currents, where present, must be parallel to the magnetic field,
\begin{equation}
	\frac{4\pi\m{J}}{c} = T'(P) \m{B} \ .
	\label{current}
\end{equation}

We require the magnetospheric toroidal field to be confined within a magnetic surface near the equator, while near the poles, where the field lines extend to very large distances, the field is current-free. This ensures smooth matching with a vacuum field at sufficiently large distances (typically 10 stellar radii). Our magnetosphere model is scalable, i.e.\ it does not depend on the overall amplitude of the magnetic field, but only on the functional relation between $P$ and $T$ and their relative amplitudes.

\subsection{Matching the interior to the magnetosphere}\label{section_boundary}
At each time step, the poloidal function $P(R_\star,\theta)$ is calculated from the radial component of the magnetic field, while the toroidal function $T(R_\star,\theta)$ is derived from the azimuthal component. In the magnetosphere, $T$ and $P$ must be functions of one another, which is not necessarily satisfied by the interior solution, where the Hall term creates deviations from such a functional relation.

The fact that $T$ is a function of $P$ implies that the solutions of the Grad--Shafranov equation must satisfy a certain symmetry, in the sense that $T$ must have the same value along a field line defined by some $P$, including at its footprints at the surface, which we can label as $\theta_1$ and $\theta_2$, so that $P(\theta_1) = P(\theta_2)$. A general (unconstrained) $T$ can then be written as the sum of a symmetric part $T_{\rm S}(\theta_1) = T_{\rm S}(\theta_2)$ and an antisymmetric part $T_{\rm A}(\theta_1) = - T_{\rm A}(\theta_2)$. The antisymmetric part cannot propagate into the magnetosphere, and is reflected back into the interior \citep[see the discussion in][]{2017MNRAS.472.3914A}. Such behavior has been observed in ideal MHD simulations of the propagation of internal torsional oscillations \citep{2014MNRAS.443.1416G}.

To address this problem, in \cite{2017MNRAS.472.3914A}, we specified a particular functional form for $T(P)$ and determined the free parameters that fitted best the values at the surface. Here, we now improve this method by allowing for a more general form of $T(P)$ by decomposing the possibly multivalued toroidal function into its symmetric and antisymmetric parts (with respect to $P$). Then, the symmetric part is used as $T(P)$ while the antisymmetric part is set to zero. Effectively, this procedure allows only the symmetric part to propagate into the magnetosphere, while the antisymmetric part must be reflected back into the interior.  

For practical purposes, we eliminate small perturbations by setting a cut-off value for $T$ (typically set at the level of $0.1\%$ of the maximum toroidal amplitude at the surface), below which we set it to zero. This cut-off value corresponds to a critical $P_{\rm c}$ below which there is no toroidal field, confining the currents into a finite region close to the neutron star.

The resulting function $T(P)$ has no fixed form and can evolve over time. In particular, it is possible to have a situation where it has a maximum somewhere in the interval $P_{\rm c} < P < P_{\rm max}$, with $P_{\rm max}$ being the maximum value of the poloidal function at the surface. In this case, $T'(P)=0$ at that point, implying zero current at the corresponding magnetic surface (as follows from equation \ref{current}). Therefore, the new generalized method allows for current reversals within the toroidal region. These reversals may happen multiple times if $T(P)$ has multiple extrema. Although this effect by itself should not cause any problems in the computation of the Grad--Shafranov equation, its implications for the stability of the resulting configuration are unclear.

Using the magnetospheric solution for $P$, we can calculate the resulting meridional component of the magnetic field ($B_\theta$) at the surface. Our more general force-free matching condition allows currents to flow through the surface, thus permitting the transfer of energy, helicity and twist between the interior and exterior.

\subsection{Initial magnetic field}\label{section_initial}
We use an initial magnetic field configuration of the form described in \cite{2017MNRAS.472.3914A}. In the interior, the poloidal component consists of a dipolar field constructed analytically for a non-barotropic background \citep{2013MNRAS.433.2445A}. To this, we superimpose a toroidal component confined within the magnetic surface defined by the critical field line $P = P_{\rm c}$,
	\begin{equation}
		T(P) \propto \left\{
		\begin{aligned}
			& (P-P_{\rm c})^2 & \mbox{for} \ P \geqslant P_{\rm c} \ , \\
			& 0 & \mbox{for} \ P < P_{\rm c} \ .
		\end{aligned}
		\right.
		\label{T_of_P}
	\end{equation}
The quadratic form ensures smoothness of the currents at the toroidal field boundary at the start of the simulations. $P_{\rm c}$ can take values in the interval from $0$ (corresponding to the pole) up to $P_{\rm max}$ (initially at the equator). We typically take $P_{\rm c} = P_{\rm max}/2$, so that the toroidal field already extends into the magnetosphere at the start of the simulation. The initial exterior field is computed as a solution of the non-linear Grad--Shafranov equation using this $T(P)$. The starting configuration contains a discontinuity in $B_\theta$ at the surface (but not in $B_r$ and $B_\phi$), which results in a surface current. As the internal field evolves, the function $T(P)$ adapts to the interior and the surface current is rapidly redistributed in a transient phase lasting a few tens of time steps, smoothing out discontinuities.

In this paper, we use the magnetic field strength at the pole $B_{\rm pole} \equiv B_r(R_\star,0)$ in order to define the initial poloidal field amplitude, while the corresponding starting toroidal field amplitude is defined by its largest value $B_{\phi,\rm max}$. We employ the same notation and dimensionless units as in our previous work \citep[see Table 1 in][]{2016MNRAS.462.1894A}. We typically use an evenly spaced angular grid of 100 points, while the radial grid has 50 evenly spaced points in the crust and 100 unevenly spaced points in the magnetosphere (distributed as a geometric series with a higher concentration near the surface).

% Sections 3 & 4: Results & Observational implications.
% Specify root document.
% !TEX root = paper.tex

\section{Results}\label{section_results}

\subsection{Sample evolution}
\begin{figure}
	\centerline{\includegraphics[width=0.48\textwidth]{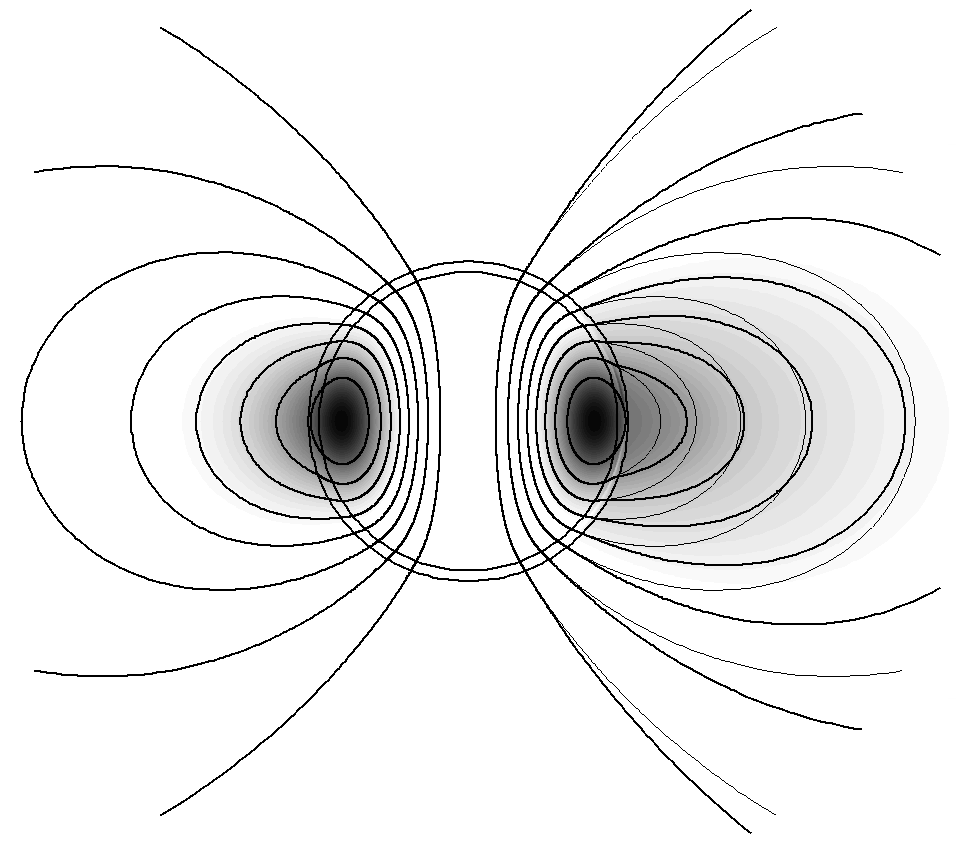}}
	\caption{Initial and final magnetic field structures for a sample run with poloidal and toroidal components of strength $10^{14}\,{\rm G}$. The configuration at the start (at $t=0$) is shown on the left hemisphere and the final configuration at the end of the evolution (at $t=2150\,$yr) is shown on the right hemisphere. For reference, the initial field configuration is also shown in thin lines in the background on the right. The crust--core boundary and the stellar surface are indicated by two circles. The gray scale represents the intensity of the toroidal function $T$ (related to $B_\phi$ through equation \ref{field}), from white (no field) to black (strongest).}
	\label{fig_snapshot}
\end{figure}

\begin{figure*}
	\centerline{\includegraphics[width=1\textwidth]{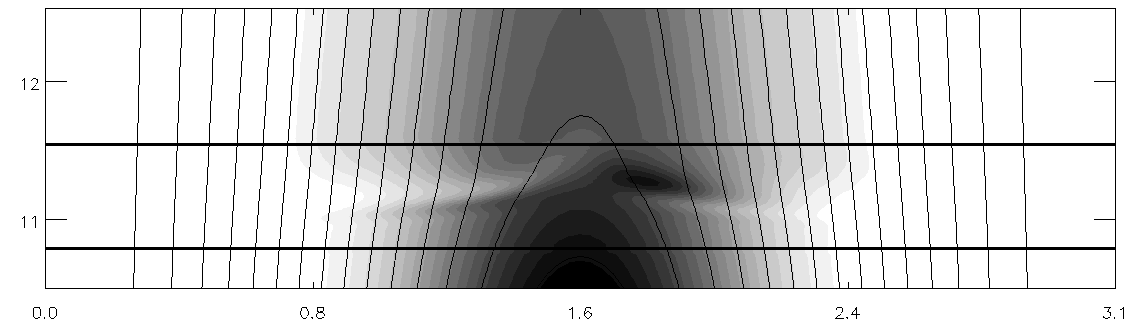}}
	\caption{Detail of the field structure in the crust and near the surface for the last snapshot of the simulation shown in Fig.~\ref{fig_snapshot} (at $t=2150\,$yr). The plot is shown as a function of the angle $\theta$ (in radians, horizontal axis) and radial distance (in km, vertical axis). The crust is located between the two thick horizontal lines --- the lower line corresponds to the crust--core boundary (at $\approx 10.8\,$km) and the upper line to the stellar surface (at $11.6\,$km). As in Fig.~\ref{fig_snapshot}, the gray scale represents the intensity of the toroidal function $T$.}
	\label{fig_snapshot_alt}
\end{figure*}

In Fig.~\ref{fig_snapshot}, we show the initial and final magnetic fields in a reference simulation for a starting configuration of the form described in \S\ref{section_initial}, with poloidal and toroidal fields of strength $10^{14}\,$G. The star is shown as a circle, and the initial and final configurations are then shown on the left and right hemispheres, respectively. The inner circle indicates the crust--core boundary. During the evolution, the toroidal amplitude near the equator remains more or less constant or decreases, but it increases towards the border of the confining surface, resulting in a gradual inflation of the force-free region and the poloidal field lines. This proceeds until a critical point is reached, beyond which no force-free solutions with connected field lines exist, implying that some other process (such as a reconnection event on a dynamical time-scale) must take place. For the case shown here, the magnetosphere reaches this point in $\sim 2150\,$yr. In Fig.~\ref{fig_snapshot_alt}, the same final magnetic field configuration is shown in a Cartesian projection as a function of the colatitude $\theta$ (horizontal axis) and the radius (vertical axis) in order to reveal more detail. The effect of the Hall term on the crustal field is now evident, where a quadrupolar (antisymmetric) component is growing, while the core field does not evolve. Throughout the evolution, the function $T(P)$ in the magnetosphere must remain single-valued, and any waves generated in the crust through departure from this constraint (i.e.\ different values of $T$ connected by the same poloidal field line defined by some $P$) are reflected back at the surface (as discused in \S\ref{section_boundary}). Our results are qualitatively in line with those presented in \cite{2017MNRAS.472.3914A}, although we have now allowed for a considerably larger degree of freedom in the relation $T(P)$ by removing constraints on its functional form.

\begin{figure}
	\centerline{\includegraphics[width=0.5\textwidth]{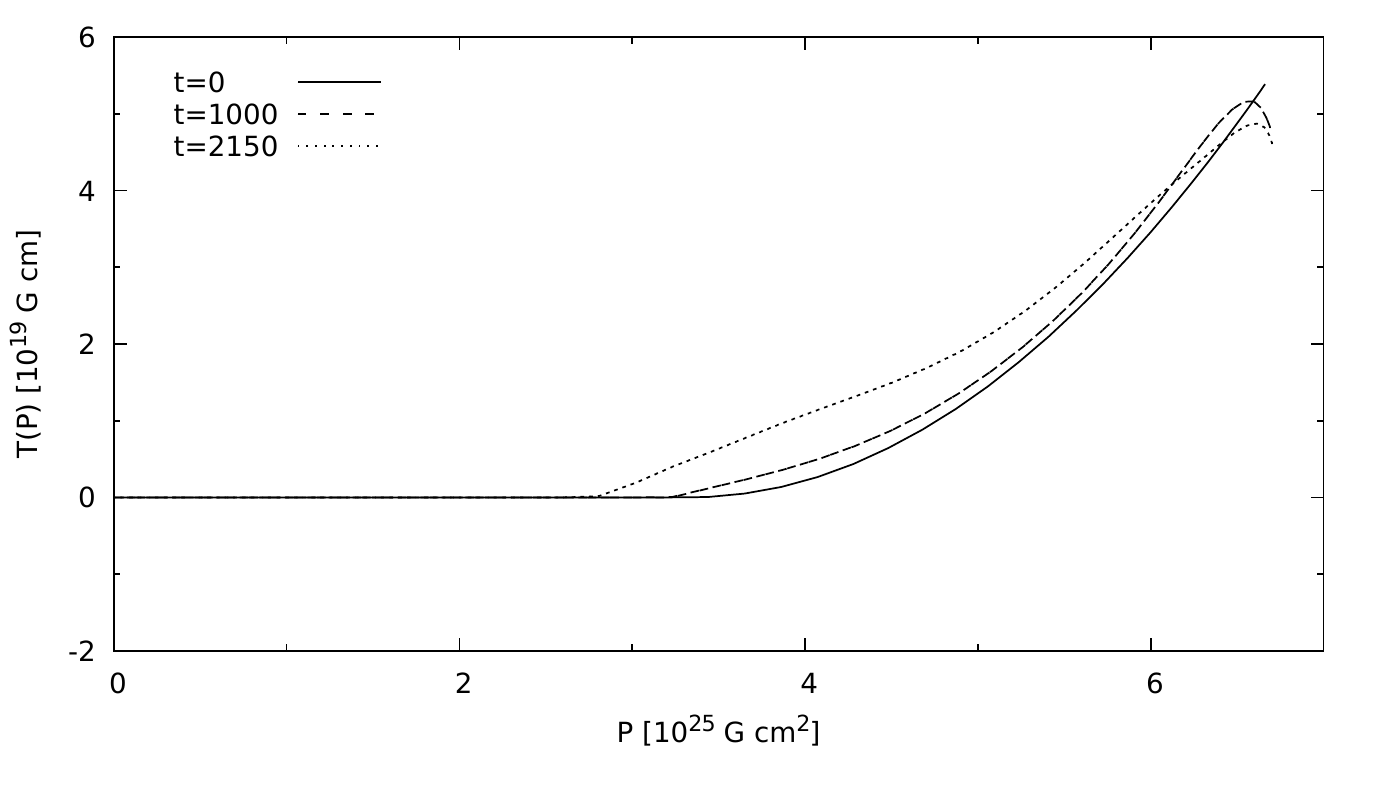}}
	\centerline{\includegraphics[width=0.5\textwidth]{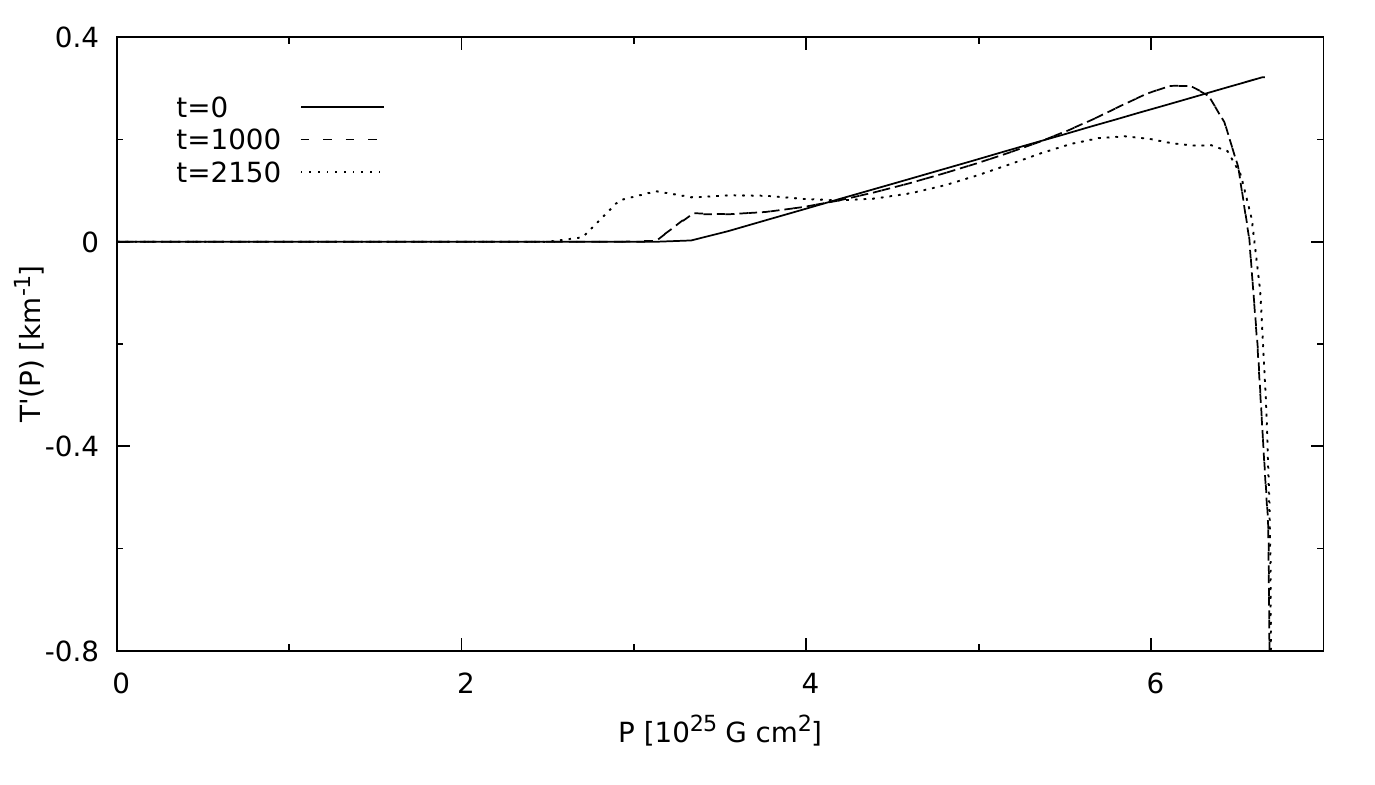}}
	\caption{Top: Snapshots of the toroidal function $T(P)$ for the magnetosphere as a function of $P$ for the same model as in Figs.~\ref{fig_snapshot} and \ref{fig_snapshot_alt}. Note that the units of $P$ and $T$ differ by a factor of length. Bottom: Snapshots for the resulting derivative $T'(P)$. A change of sign in $T'(P)$ implies reversal in the direction of currents.}
	\label{fig_bphi}
\end{figure}

\begin{figure}
	\centerline{\includegraphics[width=0.5\textwidth]{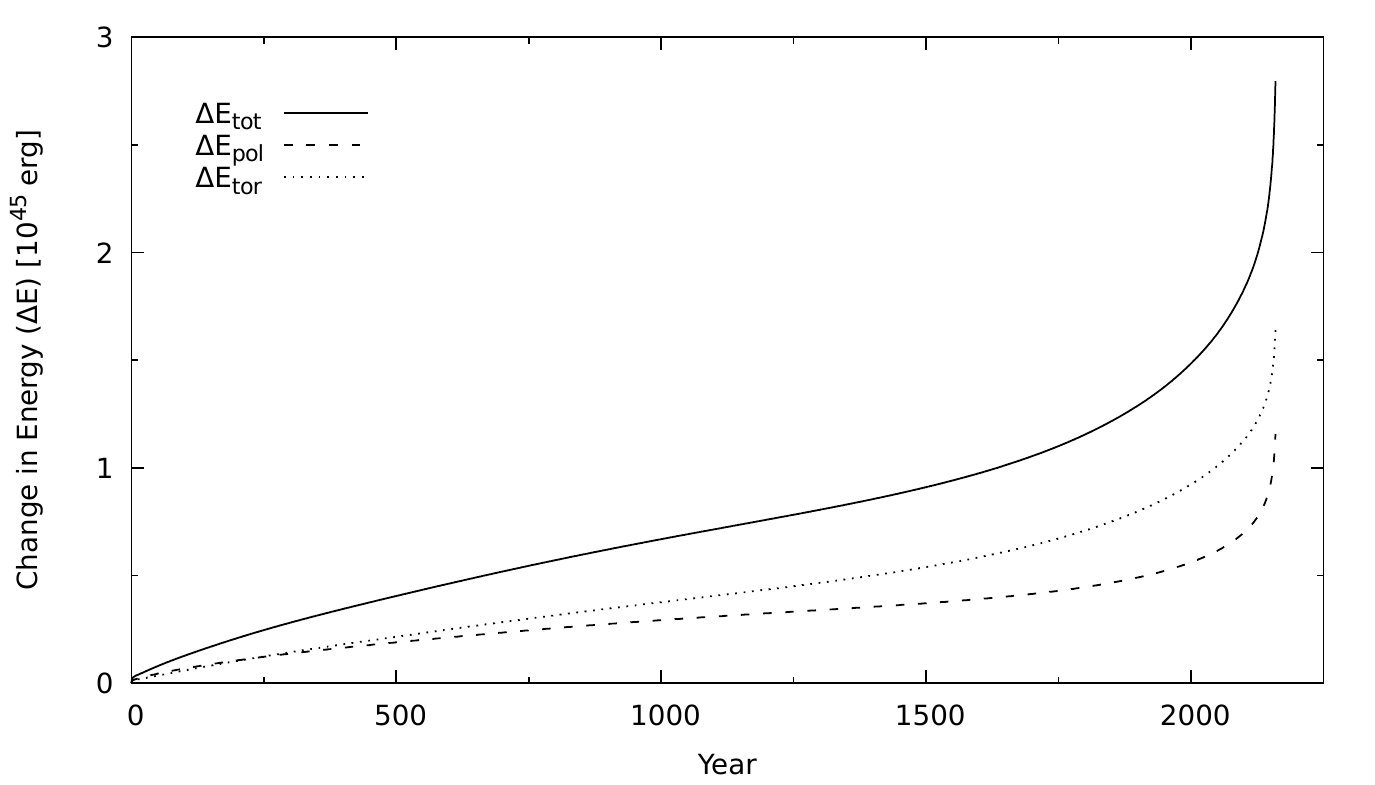}}
	\centerline{\includegraphics[width=0.5\textwidth]{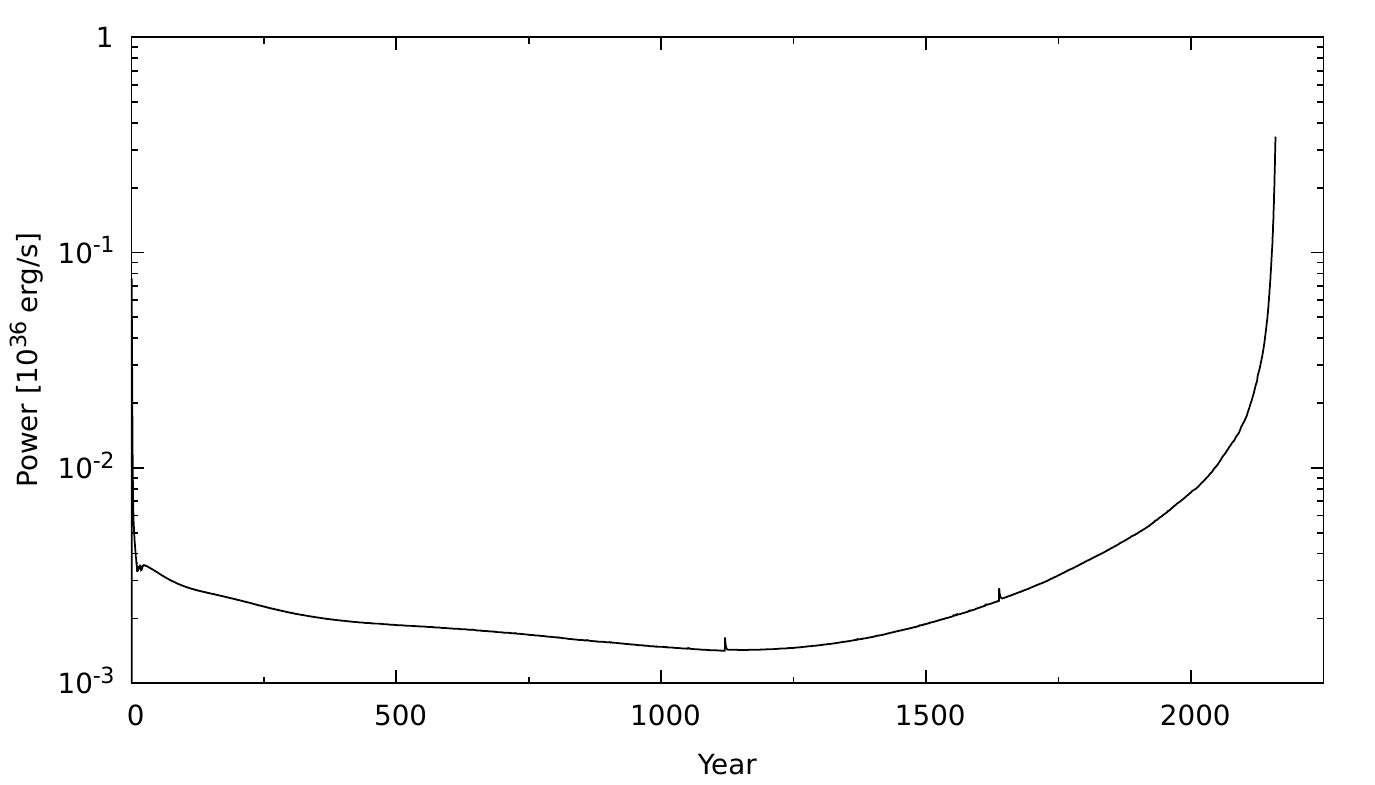}}
	\caption{Top: Evolution of the energy stored in the magnetosphere. We show the total magnetic energy ($E_{\rm tot}$), the energy of the poloidal component ($E_{\rm pol}$), and the energy of the toroidal component ($E_{\rm tor}$). As the toroidal energy is substantially lower than the poloidal energy, in order to reveal detail, we plot the changes in the energies relative to their starting values ($\Delta E = E - E_0$), which in this case are: $E_{\rm tot,0} \approx 1.38 \times 10^{45}\,$erg, $E_{\rm pol,0} \approx 1.30 \times 10^{45}\,$erg and $E_{\rm tor,0} \approx 8.22 \times 10^{43}\,$erg. Bottom: Total power input into the magnetosphere ($dE_{\rm tot}/dt$) as a function of time.}
	\label{fig_energy}
\end{figure}

Snapshots of the evolution of $T(P)$ for the magnetosphere are shown on the top panel of Fig.~\ref{fig_bphi}. Note that the equatorial torus containing currents widens over time, and the maximum value slightly decreases. Since the derivative $T'(P)$ relates the current to the magnetic field (through equation \ref{current}), the existence of a local maximum or minimum of $T(P)$ marks the transition from a region with currents circulating along the magnetic field to a region with counter-flowing currents. The lower panel of Fig.~\ref{fig_bphi} shows $T'(P)$, where we clearly see the change of sign in a narrow equatorial ring. Another important detail is that $T'(P)$ sets the inverse length scale of the dissipation of the magnetic field in the nearly force-free region just below the stellar surface. Defining $L^{-1} \equiv T'(P)$, and using equation (\ref{current}), the corresponding Joule heating rate becomes 
\begin{equation}
	Q_J = \frac{J^2}{\sigma} = \frac{\eta}{4\pi} B^2 L^{-2} \ ,
	\label{Joule_heating}
\end{equation}
where $\eta = c^2/4 \pi \sigma$ is the magnetic diffusivity. From Fig.~\ref{fig_bphi}, we can see that $L\approx 10\,$km (i.e.\ $T' \approx 0.1\,$km$^{-1}$) in most of the region where currents exist. This is a consequence of having imposed similar strengths for the toroidal and poloidal fields, and for the size of the neutron star being $R_\star \approx 10\,$km. However, there are localized regions, especially when we approach the critical point, where $L\approx 1\,$km. This has important implications as we will discuss in the next section.

\begin{figure*}
	\centerline{\includegraphics[width=0.5\textwidth]{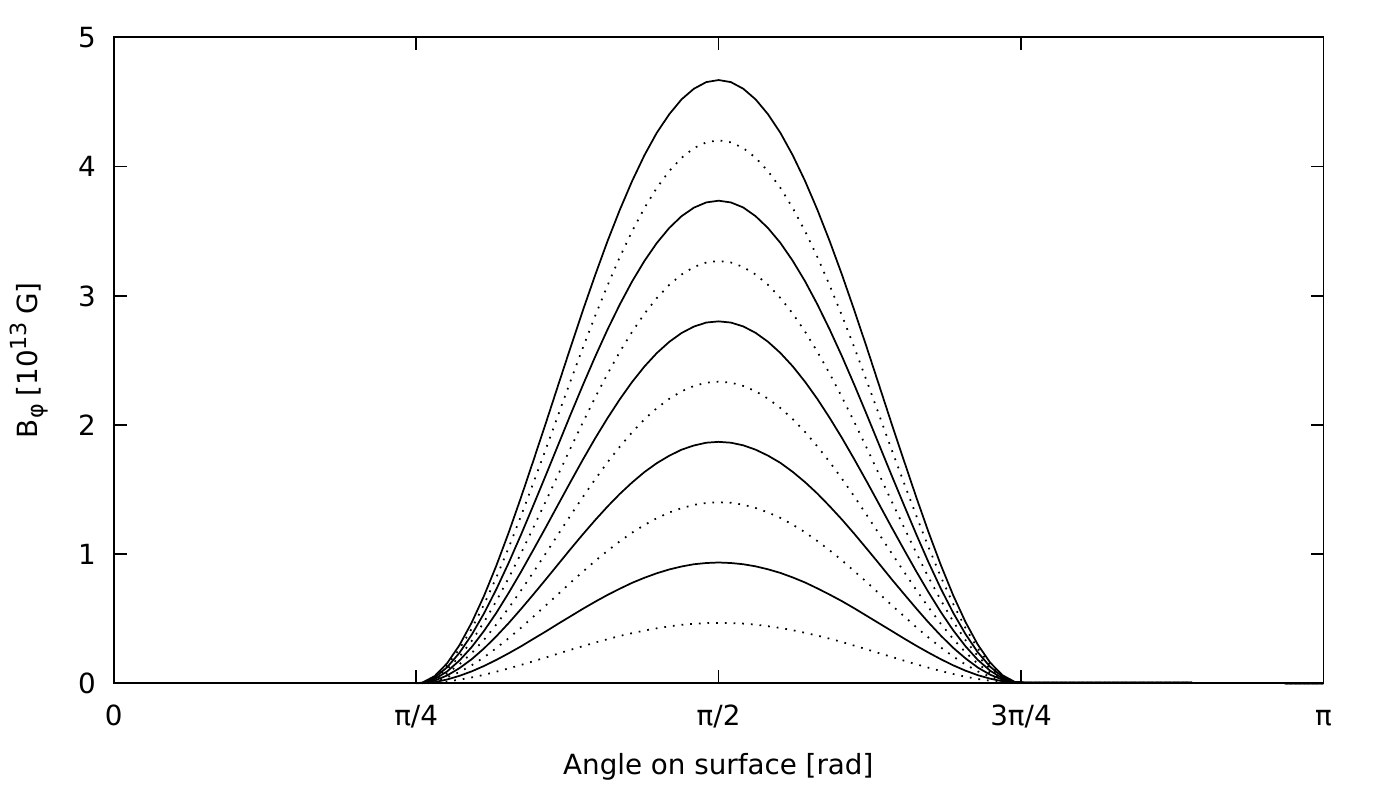}
		\includegraphics[width=0.5\textwidth]{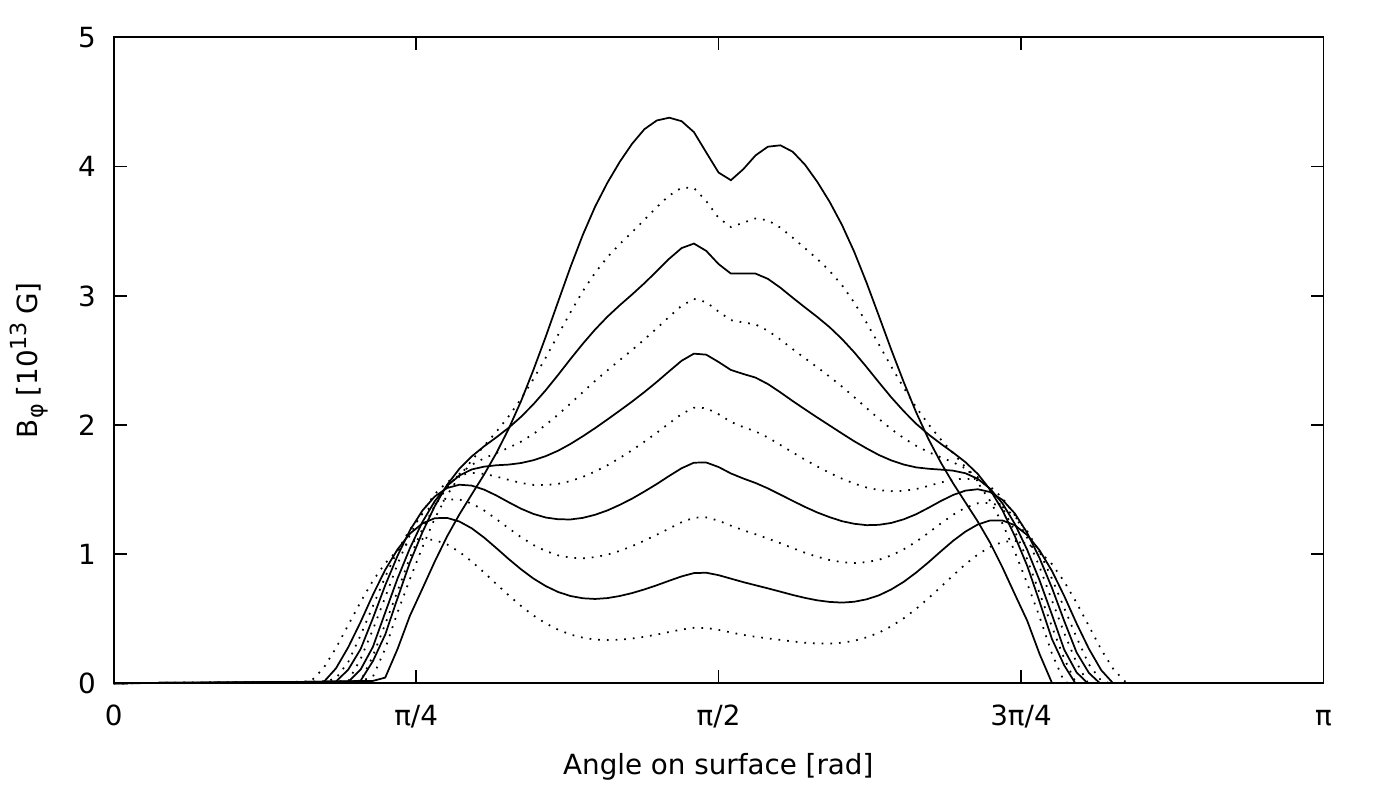}}
	\caption{$B_\phi$ at the stellar surface as a function of angle (in radians) for various toroidal field amplitudes, at the start of the simulations (left) and at the end of the simulations (right). The initial toroidal field is of the form defined by equation (\ref{T_of_P}). In these runs, the poloidal amplitude is fixed at $B_{\rm pol} = 10^{14}\,$G at the pole, and the maximum value of $B_\phi$ varies from $10^{13}\,$G up to $10^{14}\,$G, in increments of $10^{13}\,$G. Lines are shown in alternating dotted and solid lines for clarity. (Dotted lines correspond to odd amplitudes --- in units of $10^{13}\,$G --- and solid lines, to even amplitudes.)}
	\label{fig_bphi_all}
\end{figure*}

\begin{figure}
	\centerline{\includegraphics[width=0.5\textwidth]{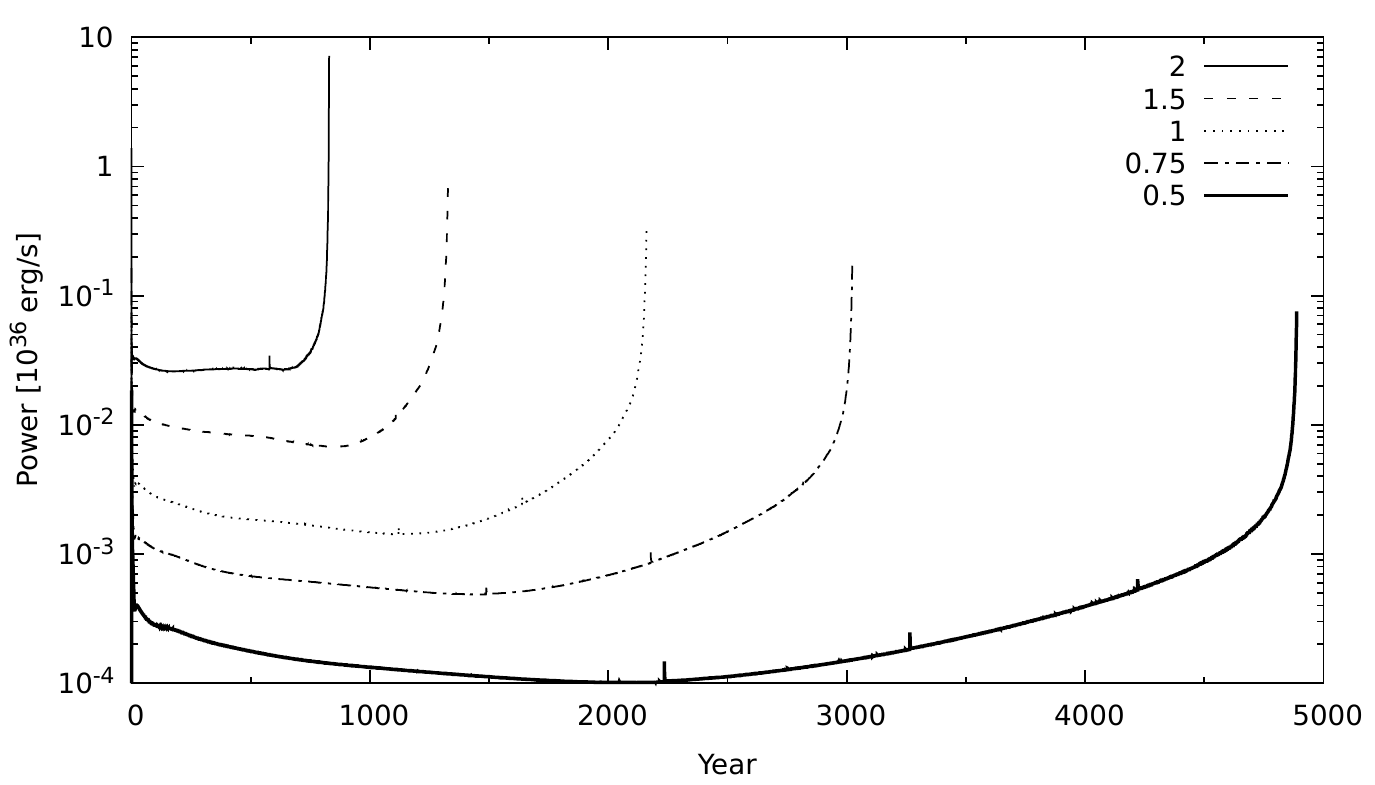}}
	\caption{Evolution of the power input into the magnetosphere for various models of different magnetic field strength. Here, the ratio of the poloidal and toroidal components is maintained the same (in this case unity), while the overall field amplitude is varied. The labels of the curves indicate the field strengths in units of $10^{14}\,$G. In all cases, the energy increases sharply near the critical point.}
	\label{fig_power_all}
\end{figure}

\begin{figure}
	\centerline{\includegraphics[width=0.5\textwidth]{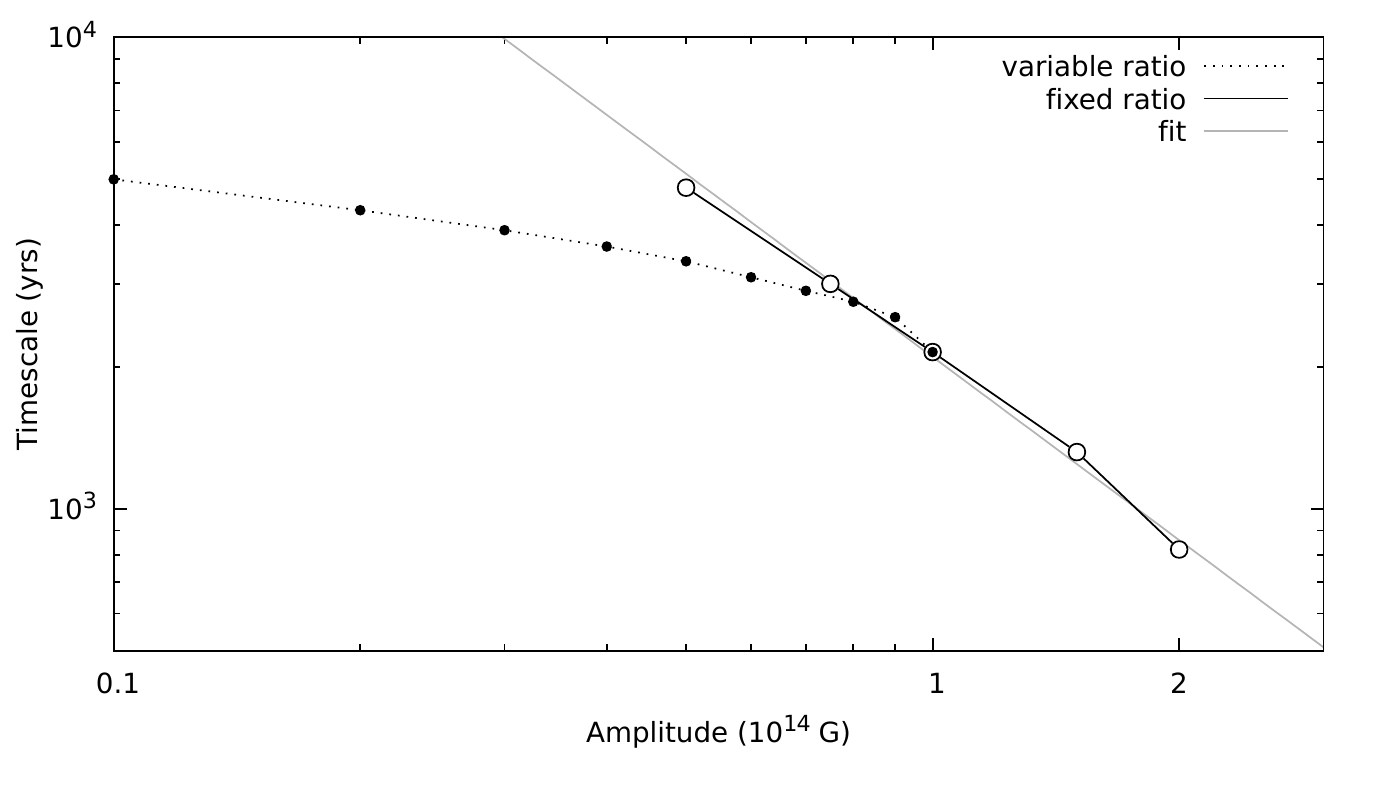}}
	\caption{Scaling of the evolution time-scale (critical time) as a function of the amplitude of the toroidal field. The black dots (dotted line) correspond to runs where the poloidal field amplitude is fixed at $10^{14}\,{\rm G}$ and the toroidal field amplitude is varied from $10^{13}\,{\rm G}$ to $10^{14}\,{\rm G}$ (as in Fig.~\ref{fig_bphi_all}). The white circles (solid line) correspond to the case where the poloidal and toroidal fields have equal amplitudes, and are varied simultaneously, i.e.\ their ratio is maintained fixed (as in Fig.~\ref{fig_power_all}). The gray line in the background is a fit to the latter case, which exhibits a more linear dependence on the log--log scale.}
	\label{fig_timescale}
\end{figure}

The evolution of the total magnetic energy stored in the entire magnetosphere (all the way to infinity) is shown on the top panel in Fig.~\ref{fig_energy}. Near the critical point the energy (for both the poloidal and toroidal components) increases rapidly. Note that, while the initial toroidal energy is rather low ($E_{\rm tor,0} \approx 8.22 \times 10^{43}\,$erg) compared to the poloidal energy ($E_{\rm pol,0} \approx 1.30 \times 10^{45}\,$erg), by the end of the simulation the toroidal energy has increased by a larger amount than the poloidal energy. If we were to compare the percentage of this increase relative to the initial poloidal and toroidal energies, we find that while the poloidal energy increases by a meager $\sim 10\%$ (with respect to $E_{\rm pol,0}$), the toroidal energy increases by nearly $\sim 200\%$ (with respect to $E_{\rm tor,0}$). The corresponding power input into the magnetosphere is shown on the bottom panel of Fig.~\ref{fig_energy}, and illustrates the rapid gain of energy in the final stages of the evolution. In other words, while the field evolution proceeds gradually for most of the simulation, near the critical point it progresses substantially faster. Interestingly, we find that the order of magnitude of the rate at which energy is transferred from the interior to the magnetosphere, driven only by the Hall drift in the crust, is of the order of the quiescence luminosities of magnetars ($10^{33}-10^{34}\,$erg/s), reaching a maximum peak of $\gtrsim 10^{35}\,$erg/s, similar to those observed during magnetar outbursts. It remains to be studied what fraction of this power can actually be released as electromagnetic radiation, and what part would get stored in the magnetosphere until it undergoes a global reconfiguration.

\subsection{Dependence on the magnetic field amplitude}
We next consider the dependence on the amplitudes of the poloidal and toroidal fields. In Fig.~\ref{fig_bphi_all}, we show the initial and final surface profiles of $B_\phi$ for various cases. We take an initial magnetic field configuration of the same form as in the previous section, and vary the toroidal field amplitude from $10^{13}\,$G up to $10^{14}\,$G, while maintaining the poloidal field amplitude fixed at $10^{14}\,$G. In all cases, the critical point appears to be reached when the border of the toroidal region shows a steep gradient (which corresponds to a large radial current). The multi-peaked form of the azimuthal component implies that multiple domains form in the vicinity of the equator where currents reverse direction (as $T'$ changes sign).

In Fig.~\ref{fig_power_all}, we show the time dependence of the power input into the magnetosphere keeping fixed the ratio of the poloidal and toroidal components (in this case unity), but varying the overall field amplitude from $5\times 10^{13}\,$G to $2\times 10^{14}\,$G. Thus, structurally, the initial magnetic field is the same and only the overall amplitude changes. As in Fig.~\ref{fig_energy}, we note that while throughout most of the evolution the power input is relatively low and constant (proportional to the field amplitude), near the respective critical points it surges by several orders of magnitude to $10^{35} - 10^{36}\,$erg/s.

The critical time (i.e.\ the time-scale to reach the critical point) decreases monotonically with increasing field amplitude as shown in Fig.~\ref{fig_timescale}. Here, we show the time-scales for the two cases discussed above: the variable ratio case of fixed poloidal and variable toroidal amplitude (dotted line), and the fixed ratio case, where the poloidal and toroidal amplitudes are changed simultaneously (solid line). The latter case is well approximated by a power law (shown in gray)
\begin{equation}
	t_{\rm c} \approx 2.10 \times 10^{3} B_{14}^{-1.29} \, {\rm yr} \ ,
	\label{timescale}
\end{equation}
where $B_{14}$ is the field amplitude in units of $10^{14}\,$G.

\section{Observational implications: surface temperatures of magnetars}\label{section_temperature}
An important implication of the existence of a long-lived force-free magnetosphere is the presence of currents flowing through the outermost $\sim 100\,$m of the neutron star (the envelope), where Ohmic dissipation may be more effective. Due to the very different thermal relaxation time-scales of the envelope and the crust,  both regions cannot be followed simultaneously  in cooling simulations. The usual approach is to employ a phenomenological fit that relates the temperature at the bottom of the envelope $T_{\rm b}$ with the surface temperature $T_{\rm s}$, in order to implement boundary conditions at the base of the envelope, typically at $\rho=10^{10}\,$g/cm$^3$. Examples of such $T_{\rm b}(T_{\rm s})$ relations for magnetized envelopes can be found in \cite{2001A&A...374..213P}, \cite{2007Ap&SS.308..353P} and \cite{Pons:2009}. We refer the reader to subsection 5.1 of the recent review by \cite{2015SSRv..191..239P} for a detailed discussion of blanketing envelopes and the calculation of transport properties under typical magnetar conditions.

\begin{figure}
	\centerline{\includegraphics[width=0.5\textwidth]{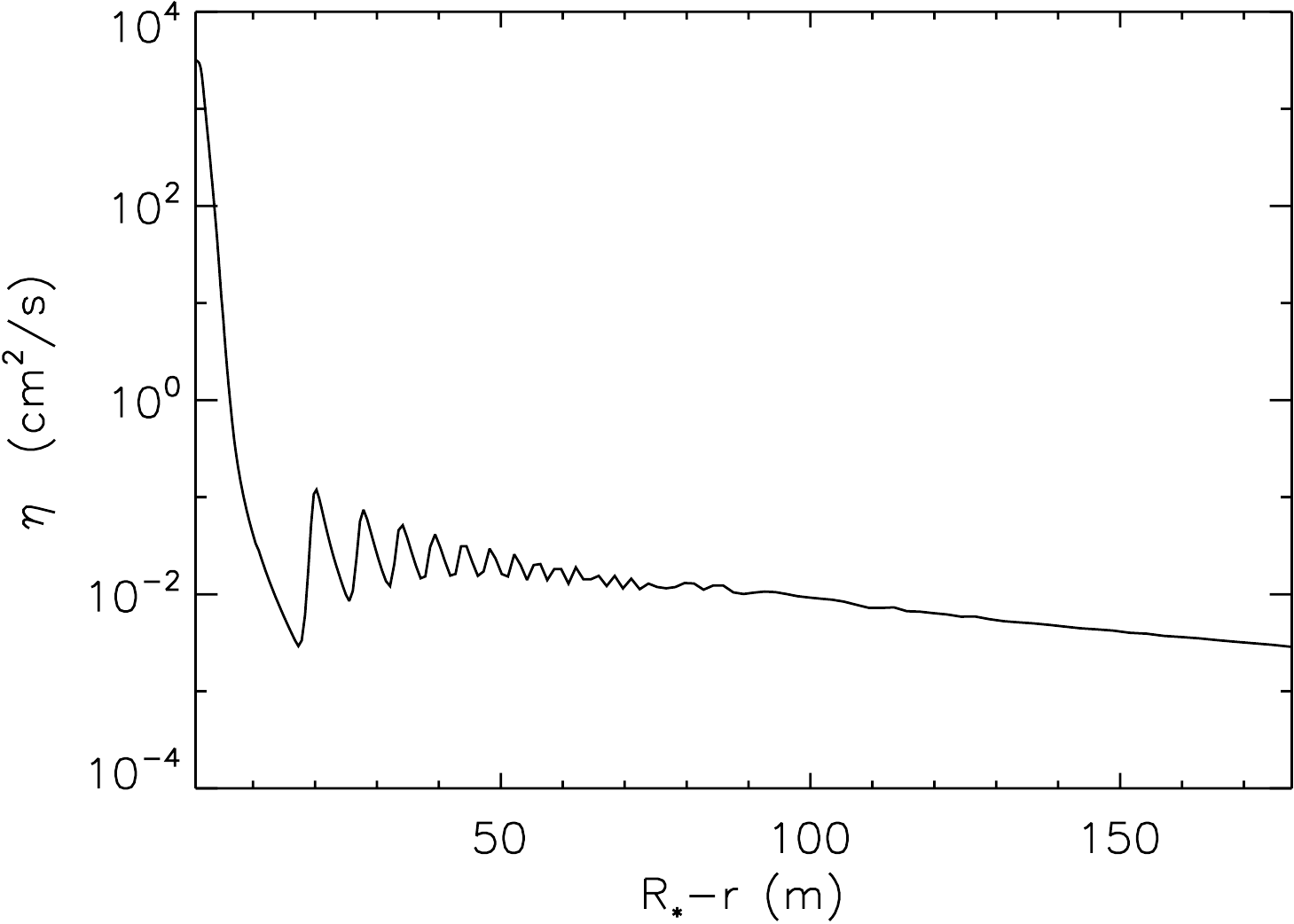}}
	\caption{Magnetic diffusivity $\eta$ as a function of depth below the surface ($R_\star - r$).}
	\label{fig_eta}
\end{figure}

In Fig.~\ref{fig_eta}, we show a profile of the magnetic diffusivity $\eta$ in a neutron star envelope. In this particular case, we have adopted $T_{\rm b}=2\times 10^8\,$K and $B =10^{14}\,$G. The spikes are due to quantizing effects, when successive Landau levels are being filled \citep[see][]{2015SSRv..191..239P}\footnote{Fortran routines for these calculations are available at: \\ {\tt http://www.ioffe.ru/astro/conduct/}}. As discussed above, the Joule heating rate can be calculated from equation (\ref{Joule_heating}), where $L$ typically varies in the range $1-10\,$km. We also note that the Ohmic dissipation time-scale is given by $\tau_{\rm Ohm} = L^2 / \eta$, which can be as short as a month in the last meter below the surface, but is of the order of years for most of the envelope. Thus, there is a crucial difference with respect to a vacuum (current-free) boundary condition: there will be a significant release of heat in the envelope that can be efficiently transported outwards, resulting in an increase of the star's surface temperature. More importantly, this can be maintained on time-scales of years (or longer, since the interior evolution may maintain the current system). The effect on the surface temperature can be estimated simply by assuming that all heat released in a volume of area $S$ and thickness $\Delta r$ is radiated as blackbody radiation. Thus, using equation (\ref{Joule_heating}), we have
\begin{equation}
	S \Delta r \frac{\eta}{4\pi} B^2 L^{-2}  = S \sigma T_{\rm eff}^4 \ ,
\end{equation}
where $\sigma$ here is the Stefan--Boltzmann constant, so that
\begin{equation}
	T_{\rm eff} \approx 0.3\,{\rm keV}
	\left[\frac{\Delta r}{1\,{\rm m}}\right]^{\frac{1}{4}}
	\left[\frac{\eta}{10^3\,{\rm cm}^2/{\rm s}}\right]^{\frac{1}{4}}
	\left[\frac{B}{10^{14}\,{\rm G}}\right]^{\frac{1}{2}}
	\left[\frac{1\,{\rm km}}{L}\right]^{\frac{1}{2}} .
\end{equation}

\begin{figure}
	\centerline{\includegraphics[width=0.5\textwidth]{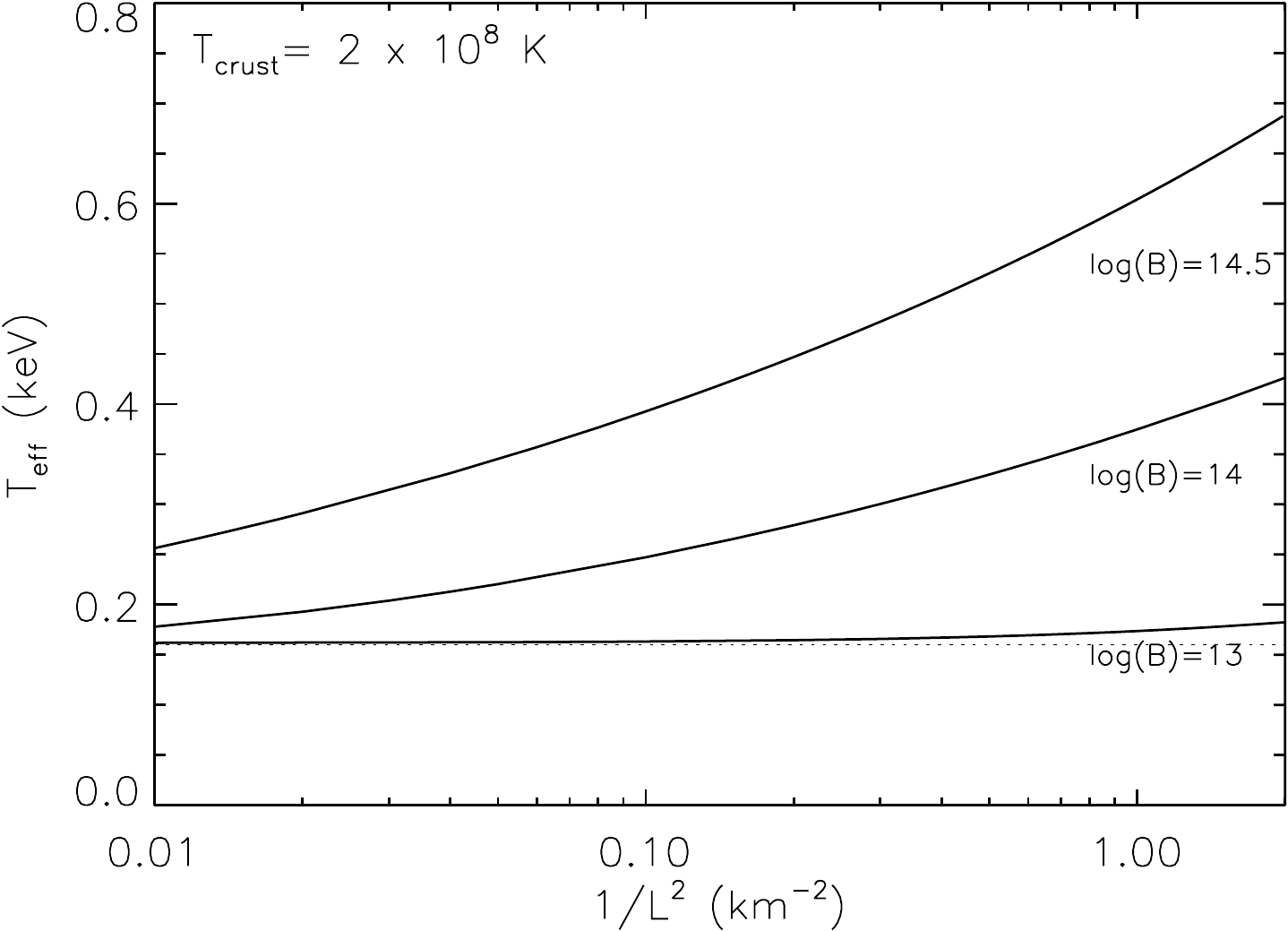}}
	\caption{Temperature as a function of length scale $L$ for various magnetic field amplitudes.}
	\label{fig_temperature}
\end{figure}

To better quantify this effect, we have recalculated 2D envelope models \citep{Pons:2009,2014MNRAS.442.3484K,2015SSRv..191..239P}, but including the effect of the heat released ($Q_J$). We assumed a typical temperature of $T_{\rm b} = 2\times 10^8\,$K at the base of the envelope and varied $B_{\rm pole}$ and the parameter $L$. The results are shown in Fig.~\ref{fig_temperature}, where we compare the effective temperature  for three cases with $B_{\rm pole} = 10^{13}$, $10^{14}$ and $3.16 \times 10^{14}\,$G. The dotted line refers to the result in the absence of currents ($Q_J=0$). For the expected range of $L$ when a force-free exterior solution is allowed, temperatures of neutron stars with relatively weak fields ($B=10^{13}\,$G) are barely affected, due to the $B^2$ dependence of $Q_J$. However, under magnetar conditions, the typical surface temperatures can be raised from $0.1\,$keV to $0.3-0.6\,$keV, in good agreement with observations \citep{2017MNRAS.471.1819C}. Therefore, we conclude that the observed flux and temperature evolution during magnetar outbursts is consistent with the expected heat release by a current system extending from the crust to the magnetosphere. This heat released would be concentrated in the outermost few meters, which turns out to be very effective in increasing the surface temperature. Obviously, our 2D models are limited by axial symmetry, our hot spots are actually hot rings,  and we do not allow for currents near the pole, which will concentrate this effect in a smaller area. In a realistic 3D case, one may expect that the magnetic field evolution driven by the Hall drift in the crust would occasionally result in a flare, creating a coronal-like magnetic loop affecting a typical area of $1-10\,$km$^2$ which may be maintaned for a relatively long time-scale (of the order of years). A more detailed quantitative study requires 3D simulations, which are not yet available.

\section{Conclusions}\label{section_conclusions}
In this paper, we have continued and extended our previous work on force-free magnetospheres, focusing on the effect of the coupling with the internal magneto-thermal evolution. The main technical improvement here is that the construction of the function $T(P)$ for the magnetosphere at the stellar surface has been generalized. While in \cite{2017MNRAS.472.3914A} we carried out a quadratic best fit to determine the relation $T(P)$, here we separate it into symmetric and antisymmetric parts (with respect to $P$), allowing the symmetric part to propagate into the magnetosphere, while reflecting the antisymmetric part back into the interior (as required by the force-free condition). Thus, the toroidal function is allowed to evolve freely (though consistently with the interior) without any imposed prescriptions on its particular form. This larger freedom in the choice of $T(P)$ allows for the formation of regions in the magnetosphere with current reversal, whenever $T'(P)$ changes sign. This also allows us to handle  higher values for the toroidal to poloidal field ratio. 

We find that, qualitatively, the field evolution follows the same stages as in \cite{2017MNRAS.472.3914A}: for most part of the evolution, the toroidal region in the magnetosphere gradually grows, while the interior field evolves under the dominant Hall term. The growth of the magnetospheric currents proceeds until a critical point is reached beyond which force-free solutions for the magnetosphere (given as solutions of the Grad--Shafranov equation) cannot be constructed, likely leading to some large-scale magnetospheric reorganization such as a burst or a flare. The energy budget available for a magnetospheric event can now be as high as several $10^{45}\,$erg (Fig.~\ref{fig_energy}).

The critical time (i.e.\ the time it takes to reach the critical point) is typically in the range of a few thousand years and is inversely related to the magnetic field amplitude (Fig.~\ref{fig_timescale} and equation \ref{timescale}). Near this critical point the power input from the interior into the magnetosphere increases by several orders of magnitude to $10^{35} - 10^{36}\,$erg/s (Fig.~\ref{fig_power_all}), which is consistent with peak luminosities during magnetar outbursts, and also suggests that some kind of precursor activity of an outburst could be potentially observed.

We also comment on an observationally relevant property of our force-free magnetosphere model: allowing currents to flow through the surface has important implications for the local temperature. In particular, strong currents passing through the last hundred meters of the surface (the envelope), especially in the last few meters where the magnetic diffusivity is orders of magnitude larger (Fig.~\ref{fig_eta}), should give rise to a considerable amount of energy being deposited very close to the stellar surface through Joule heating. We estimate that when a  magnetosphere is established, the effective surface temperature could increase locally from $\sim 0.1\,$keV to $\sim 0.3 - 0.6\,$keV (Fig.~\ref{fig_temperature}), in good agreement with observations. Therefore, a careful and detailed treatment of currents flowing through the envelope may be a key ingredient, although often overlooked, to explain the thermal properties of magnetars. More detailed calculations, particularly 3D models, therefore seem necessary.

In addition, it is conceivable that there is a threshold value below which the magnetic field is too weak for the continuous replenishment of currents in the magnetosphere. To precisely determine this value, one must consider the balance between the rate at which energy is transferred into the magnetosphere and the local dissipation rate in the last few meters of the star, coupled with the temperature evolution. Such a high resolution study has not yet been possible with present numerical cooling codes, which usually evolve only the crust and the core and consider the outer layers through a boundary condition, given the vastly different (by many orders of magnitude) thermal relaxation times. In light of our results, some effort must be put in this direction to better understand the magnetar emission properties.

\section*{Acknowledgements}
We thank Daniele Vigan\`{o} for useful comments. This work is supported in part by the Spanish MINECO/FEDER grants AYA2015-66899-C2-1-P, AYA2015-66899-C2-2-P, the grant of Generalitat Valenciana PROMETEOII-2014-069, and by the PHAROS COST action CA16214. P.~C. acknowledges the support from the Ram\'{o}n y Cajal program of the Spanish MINECO (RYC-2015-19074).

% Bibliography.
\bibliographystyle{mnras}
\bibliography{references}

% End of document.
\label{lastpage}
\end{document}